\documentclass[twocolumn]{revtex4}
\usepackage{amsmath,lscape,epsfig}

\def\beq{\begin{equation}}
\def\eeq{\end{equation}}
\def\beqa{\begin{eqnarray}}
\def\eeqa{\end{eqnarray}}
\def\ban{\begin{eqnarray*}}
\def\ean{\end{eqnarray*}}
\def\bi{\begin{itemize}}
\def\ei{\end{itemize}}

\begin{document}

\title{Particle Production within the Quark Meson Coupling Model}

\author{P.K. Panda}
\affiliation{Indian Association for the Cultivation of Science,
Jadavpur, Kolkata-700 032, India}
\affiliation{Centro de F\'{\i}sica Computacional, Department of Physics,
  University of  Coimbra,  3004-516 Coimbra, Portugal}
\author{D.P.Menezes}
\affiliation{Depto de F\'{\i}sica - CFM - Universidade Federal de Santa
Catarina  Florian\'opolis - SC - CP. 476 - CEP 88.040 - 900 - Brazil}
\author{C. Provid\^encia}
\affiliation{Centro de F\'{\i}sica Computacional, Department of Physics,
  University of  Coimbra,  3004-516 Coimbra, Portugal}

\begin{abstract}

Quark meson coupling (QMC) models can be successfully applied to the description
of compact star properties in nuclear astrophysics as well as to
nuclear matter. In the regime of hot hadronic matter very
few calculations exist using the QMC model, in
particular when applied to particle yields in heavy ion collisions.
In the present work, we identify the free energy of the bag with the effective 
mass of the baryons and  we calculate the particle production yields on a 
Au+Au collision at RHIC with the QMC model and compare them with results
obtained previously with other relativistic models. A smaller temperature for 
the fireball, T=132 MeV, is obtained due to  the smaller effective baryon 
masses  predicted by QMC. QMC was also applied to the description of 
particle yields at SPS in Pb+Pb collisions.
\end{abstract}

\maketitle
\vspace{0.50cm}
PACS number(s): {21.65.+f, 24.10.Jv, 95.30.Tg}
\vspace{0.50cm}

\section{Introduction}
The knowledge of the equation of state (EoS) of nuclear matter
under exotic conditions, including high isospin asymmetries, finite
temperatures, and a wide density range, is essential for our understanding of 
the nuclear force. It is important to impose constraints coming both from 
laboratory measurements and astrophysical on the nuclear 
models presently used to describe nuclear matter.
 
The quark-gluon plasma (QGP) phase refers to matter where quarks and gluons
are believed to be deconfined and it probably takes place at temperatures of the
order of 150 to 170 MeV. These temperatures were possible in nature only 
shortly after the Big Bang. In large  colliders around the world 
(RHIC/BNL, ALICE/CERN, GSI, etc), physicists are trying to
convert hadronic matter at sufficiently high temperatures into QGP. Possible
experiments towards this search are Au-Au collisions at RHIC/BNL
and Pb-Pb collisions at SPS/CERN, where the
hadron abundances and particle ratios are used in order to determine the 
temperature and baryonic chemical potential of the possibly present 
hadronic matter-QGP phase transition.

Recently relativistic nuclear models have been tested in the high temperature 
regime produced in these heavy ion collisions. 
In previous works these data have already been analyzed 
\cite{munzinger,munzinger2,schaffner,nosso1,nosso2} under different 
perspectives. In 
\cite{munzinger2} the authors have used a statistical model which assumes 
chemical equilibration to find the temperature and baryon chemical potential
that provide a best fit to the data obtained by the NA49 \cite{na49}
and WA97 \cite{wa97} collaborations. In this work the interaction among the 
baryons and mesons were neglected and an eigenvolume was assigned to all 
particles so that repulsive interactions among hadrons were
considered. In \cite{schaffner} the nuclear interaction was included  
through a relativistic self-consistent chiral
model of hadrons, which embodies the restoration of chiral symmetry at both
  high temperatures and densities. The results depended on the parametrizations 
used and indicated that no direct freeze-out from the restored phase was 
observed. 

In \cite{nosso1} four parametrizations of the non-linear Walecka model
\cite{sw}, namely NL3 \cite{nl3}, TM1 \cite{tm1}, GM1 and GM3 \cite{glen}, 
one model with implicit density dependence through meson field couplings, the
NL$\omega\rho$ \cite{nlwr} and two different parametrizations of a
density dependent hadronic model, the TW \cite{tw} and the DDME1
\cite{twring} were used to calculate the Au-Au collision particle yields. 
Eighteen baryons, 3 mesons to mediate the nuclear force and pions, kaons, 
$\rho$s and $K^*$s were included.
It was shown that if the light mesons, e.g. pions and kaons, are not taken in
the interaction with baryons the models do not have enough repulsion among 
hadrons and are not able to reproduce experimental data with the same quality 
as the thermal model \cite{munzinger,munzinger2} or  the  relativistic 
self-consistent chiral model used in \cite{schaffner}. 
Within the thermal model 
the  particle production fractions are reproduced with a temperature  
$T=174 \pm 7$ MeV and a baryonic chemical potential  $\mu_B=46 \pm 5$ MeV, 
while for the chiral model these quantities  are $T=155$ MeV and the baryon 
chemical potential of the order of $\mu_B=51$ MeV.
In \cite{nosso1} these numbers lie in the range $146 < T < 153$ MeV and
$46.5 < \mu_B < 62.8$ MeV. In \cite{nosso2} the parameters related to the 
coupling of the hyperons to the mesons were adjusted in accordance with the 
different hyperonic binding energies and the numbers for the freeze-out 
temperature and chemical potential with a modified GM3 \cite{glen} 
parametrization were $T=147.7$ MeV and $\mu_B=31.6$ MeV. It is worth
mentioning that all those numbers depend on the set of hyperon couplings, 
a value not well known.

In the present paper we test the behaviour of the quark meson coupling model
(QMC) in this regime of temperature and density. Within the QMC model, nuclear
matter is described as a system of nonoverlapping MIT bags which interact
through the effective scalar and vector mean fields \cite{qmc}.
Although the QMC model shares many similarities with the non-linear Walecka 
models (NLWM) \cite{sw}, it also
offers new opportunities for studying nuclear matter properties. One of the
most attractive aspects of the model is that different phases of hadronic
matter, from very low to very high baryon densities and
temperatures, can be described within the same underlying model.

In the QMC the internal structure of the nucleon is introduced explicitly and  
matter at low densities and temperatures is a system of nucleons interacting 
through meson
fields, with quarks and gluons confined within MIT bags. For matter at very
high density and/or temperature, one expects that baryons and mesons dissolve
and the entire system of quarks and gluons becomes confined within a single,
big MIT bag. In most cases in the literature,
the energy of the nucleonic MIT bag is identified with the
effective mass of the nucleon. This identification has important implications: 
at finite temperature, while in the NLWM
models the nucleon mass always decreases with temperature, in the QMC it
increases \cite{pandaqmc}.
The difference arises due to the explicit treatment 
of the internal structure of the nucleon in the QMC. When the bag is heated up,
quark-antiquark pairs are excited in the interior of the bag, increasing the
internal energy of the bag.  In the present approach, we identify the
effective mass of the nucleon with the free energy of the bag, and as 
it is shown next a direct consequence is the recovery of the behaviour 
of the NLWM for the effective mass, i.e., it decreases with the 
increase of the temperature.
This choice makes sense because we want to identify the
temperature as a state variable. Next we discuss some of the consequences of 
this choice and apply the QMC to the description of  hadron
abundances  and particle ratios in Au-Au with $s=\sqrt{130}$ GeV collisions at
RHIC/BNL and Pb-Pb collisions at SPS. 

The paper is organized as follows: in section II we give a brief review of the 
QMC model and its generalization for finite temperatures; in section
III we present some results refering the description of warm nuclear matter
within the QMC and apply the formalism to the description of particle 
production in  Au-Au $s=\sqrt{130}$ GeV collisions at RHIC/BNL
and Pb-Pb collisions at SPS.; in section IV 
we draw our final conclusions.
\section{Formalism}
In the QMC model, the nucleon in nuclear medium is assumed to be a
static spherical MIT bag in which quarks interact with the scalar and
vector fields, $\sigma$, $\delta$,  $\omega$ and $\rho$ and these
fields are treated as classical fields in the mean field
approximation\cite{qmc,pandaqmc}. The quark field
$\psi_q({\bf r},t)$ inside the bag then satisfies the Dirac equation
\begin{equation}
\Big[i{\vec \gamma}\cdot{\vec \partial}-(m_q^0-V_\sigma)-\gamma^0(V_\omega
+\frac{1}{2} \tau_{3q}V_\rho) \Big]\psi_q({\bf r},t)=0,
\end{equation}
with $ q=u,d,s,$ where $V_\sigma=g_\sigma^q\sigma_0$, 
$V_\omega=g_\omega^q\omega_0$ and
$V_\rho=g_\rho^q~\rho_{03}$
with $\sigma_0$, $\omega_0$ and $\rho_{03}$ being the classical meson
fields. $g_\sigma^q$, $g_\omega^q$ and $g_\rho^q$
are the quark meson couplings with the $\sigma$, $\omega$ and $\rho$ mesons
respectively and $m_q^0$ is the current quark mass.
The normalized ground state for a quark in the bag is given
by
\begin{equation}
\psi_q({\bf r}, t) = {\cal N}_q \exp\left(-i\epsilon_q t/R_B \right)
\left(
\begin{array}{c}
j_0\left(\frac{x_q }{R_B}\right)\\
i\beta_q \vec{\sigma} \cdot \hat r j_1\left(\frac{x_q r}{R_B}\right)
\end{array}\right)
 \frac{\chi_q}{\sqrt{4\pi}} ~,
\end{equation}
where
\begin{equation}
\epsilon_q=\Omega_q +R_B\left(g_\omega^q\, \omega_0+
\frac{1}{2} g^q_\rho \tau_z \rho_{03} \right)  ~; ~~~
\end{equation}
$$
\beta_q=\sqrt{\frac{\Omega_q-R_B\, m_q^*}{\Omega_q\, +R_B\, m_q^* }}\ ,
$$
with the normalization factor given by
\begin{equation}
{\cal N}_q^{-2} = 2R_B^3 j_0^2(x_q)\left[\Omega_q(\Omega_q-1)
+ R_B m_q^*/2 \right] \Big/ x_q^2 ~,
\end{equation}
where $\Omega_q\equiv \sqrt{x_q^2+(R_B\, m_q^*)^2}$,
$m_q^*=m_q^0-g_\sigma^q\, \sigma_0$, $R_B$ is the
bag radius of the baryon $B$,
and  $\chi_q$ is the quark spinor. The quantities $\psi_q,\, \epsilon_q,\,
\beta_q,\, {\cal N}_q,\, \Omega_q,\, m^*_q$ all depend on the baryon
considered. The bag eigenvalue, $x_q$, is determined by the
boundary condition at the bag surface
\begin{equation}
j_0(x_q)=\beta_q\, j_1(x_q)\ .
\label{bun-con}
\end{equation}
At finite temperatures, the three quarks inside the bag can be thermally 
excited to higher angular momentum states and also quark-antiquark
pairs can be created. For simplicity, we assume that the bag describing the 
nucleon continues to remain in a spherical shape with radius $R$, which is now
temperature dependent. The single-particle energies in units
of $R^{-1}$ are given as
\begin{equation}
\epsilon_{q}^{n\kappa}=\Omega_q^{n\kappa}+ 
R_B(V_\omega \pm \frac{1}{2}V_\rho ),
\end{equation}
for the quarks and
\begin{equation}
\epsilon_{\bar q}^{n\kappa}=\Omega_q^{n\kappa}- R_B(V_\omega \pm
\frac{1}{2}V_\rho ),
\end{equation}
for the anti-quarks, where the $+$ sign is for $u$ quarks and $-$ for $d$
quarks, and
\begin{equation}
\Omega^{n\kappa}_q = \sqrt{x^2_{n\kappa}+R_B^2{m^*_q}^2} .
\end{equation}
The eigenvalues $x_{n\kappa}$ for the state characterized by $n$ and $\kappa$
are determined by the boundary condition at the bag surface,
\begin{equation}
i\gamma\cdot n \psi_q^{n\kappa}=\psi_q^{n\kappa} .
\end{equation}
Thus, the quark eigenvalues $x_{n\kappa}$ become modified by the surrounding
nucleon medium at finite temperature.
The total energy from the quarks and anti-quarks at finite temperature is
\begin{equation}
E_{tot} = \sum_{q,n,\kappa}\frac{\Omega^{n\kappa}_q}{R_B}\left(f^q_{n\kappa} +
f^{\bar q}_{n\kappa}\right),
\label{Etot}
\end{equation}
where
\begin{eqnarray}
f^q_{n\kappa}& =&  \frac{1}{e^{(\Omega_{q}^{n\kappa}/R_B-\nu_q)/T}+1} \,
\nonumber\\
f^{\bar q}_{n\kappa} &=& \frac{1}{e^{(\Omega_{q}^{n\kappa}/R_B+\nu_q)/T}+1},
\label{fq}
\end{eqnarray}
with $\nu_q$ being the effective quark chemical potential, related to the
 quark chemical potential $\mu_q$ as
\begin{equation}
\nu_q=\mu_q-V_\omega-m_\tau^q \, V_\rho .
\label{nuq}
\end{equation}
The energy of a static bag describing baryons consisting of three ground 
state quarks can be expressed as
\begin{equation}
E^{\rm bag}_B=E_{tot}-\frac{Z_B}{R_B}
+\frac{4}{3}\,  \pi \, R_B^3\,  B_B\ ,
\label{ebag}
\end{equation}
where $Z_B$ is a parameter which accounts for zero-point motion
and $B_B$ is the bag constant.
The entropy of the bag is defined as
\begin{eqnarray}
{\cal S}^{\rm bag}_B &=& -\sum_{q,n,\kappa}
\Big[ f^q_{n\kappa}\ln f^q_{n\kappa}+(1-f^q_{n\kappa})\ln (1-f^q_{n\kappa})\nonumber \\
&+& \bar f^q_{n\kappa}\ln \bar f^q_{n\kappa}+(1-\bar f^q_{n\kappa})\ln (1-\bar f^q_{n\kappa})\Big] ,
\label{entropyq}
\end{eqnarray}
and the free energy for the bag is given by
\begin{equation}
F_B^{\rm bag}=E_B^{\rm bag}+T~S_B^{\rm bag}
\end{equation}
The set of parameters used in the present work is given
in Ref. \cite{cfl}. The effective mass of a nucleon bag at rest
is taken to be 
\begin{equation}
M_B^*=F_B^{\rm bag}.
\label{meff}
\end{equation}
In reference \cite{pandaqmc}, it was considered the
bag energy instead of the free energy to define the effective mass.
The equilibrium condition for the bag is then obtained by
minimizing the effective mass, $M_B^*$ with respect to the bag radius
\begin{equation}
\frac{d\, M_B^*}{d\, R_B^*} = 0\ .
\label{balance}
\end{equation}
Once the bag radius is obtained, the effective baryon mass is immediately
determined. For a given temperature $T$ and scalar field $\sigma$, the
effective quark chemical potentials, $\nu_q$, are determined
from the total number of quarks, isospin density and strangeness,
i.e.,
\begin{eqnarray}
n^j_0 &=& \sum_{q,n,\kappa}\left(f^q_{nq} - f^{\bar q}_{nq}\right)
\equiv 3,
\label{muqa} \\
n^j_3 &=& \sum_{q,n,\kappa} \, 2m_{\tau(q)}\left(f^q_{nq} -
f^{\bar q}_{nq}\right) \equiv 2 m_{\tau(j)},
\label{muqa1}\\
r^j_s &=& \sum_{q,n,\kappa} \, r_s(q)\left(f^q_{nq} -
f^{\bar q}_{nq}\right).
\label{muqa2}
\end{eqnarray}
In our calculation we consider  $j=\Lambda$.

The total energy density of baryonic matter at finite temperature $T$ and 
at finite baryon density $\rho_B$ is
\begin{eqnarray}
{\cal E} &=&\frac{2}{(2\pi)^3}\sum_{i=B}\int d^3 k \, \left[\epsilon^*\,
(f_i+\bar f_i)+
{\cal V}_{0i}(f_i-\bar f_i)\right]\nonumber\\
&+& \frac{1}{2}{m_\sigma^2}\sigma^2-\frac{1}{2}{m_\omega^2}
\omega^2-\frac{1}{2}{m_\rho^2} \rho_{03}^2,
\label{ener1}
\end{eqnarray}
where $f_i$ and $\bar f_i$ are the thermal distribution functions
for the baryons and anti-baryons,
\begin{equation}
f_B=\frac{1}{e^{(\epsilon^*-\nu_B)/T}+1} ~~{\rm and}~~
\bar f_B=\frac{1}{e^{(\epsilon^*+\nu_B)/T}+1},
\end{equation}
$\epsilon^*=(\vec k^2+{M^*_B}^2)^{1/2}$ the effective
nucleon energy,  $\nu_B=\mu_B-{\cal V}_{0B}$ the effective
baryon chemical potential  and ${\cal V}_{0B}=g_{\omega B} \omega+ I_{3B}\,
g_{\rho B} b_{03}\,$  ($I_{3B}$ is the   isospin projection of the baryon 
species $B$). The couplings of the mesons with the baryons, $g_{\omega B}$ 
and  $g_{\rho B}$, will be discussed below. 
The thermodynamic grand potential density and the free energy density are 
defined as
\begin{equation}
\Omega = {\cal F}- \sum_{i=B} \mu_i\rho_i, \quad 
{\cal F}={\cal E} - T {\cal S},
\label{therm}
\end{equation}
with the entropy density ${\cal S} = S/V$ given by
\begin{eqnarray}
{\cal S} &=& -\sum_{i=B}\frac{2}{(2\pi)^3}\int d^3 k
\Big[ f_i\ln f_i+(1-f_i)\ln (1-f_i)\nonumber \\
&+& \bar f_i\ln \bar f_i+(1-\bar f_i)\ln (1-\bar f_i)\Big] .
\label{entropy}
\end{eqnarray}
The baryon density (of each baryon species) is given by
\begin{equation}
\rho_{i}=\frac{2}{(2\pi)^3}\int d^3 k ~(f_i-\bar f_i),
\end{equation}
so that the total baryon density is $\rho=\sum_{i=B}\rho_i$.
The pressure is the negative of $\Omega$, which after an integration by parts
can be written as
\begin{eqnarray}
P&=& \frac{1}{3}\sum_{i=B}\frac{2}{(2\pi)^3}\int d^3 k
\frac{{\bf k}^2}{\epsilon^*(k)} (f_i+\bar f_i) \nonumber \\
&-& \frac{1}{2}{m_\sigma^2}\sigma^2 + \frac{1}{2}{m_\omega^2}
\omega^2+\frac{1}{2}{m_\rho^2} \rho_{03}^2.
\end{eqnarray}
From the above expression the pressure depends explicitly on the
meson mean fields $\sigma$, $\omega$ and $\rho_{03}$. It also depends
on the baryon effective mass $M^{*}_B$ which in turn also depends on  the 
sigma field 
(see Eqs.~(\ref{Etot}-\ref{balance})).
At a given temperature and for given baryon density, the effective
mass is known for given values of the meson fields, once the bag radius $R_B$
and the effective quark chemical potentials $\nu_q$  are calculated by using
Eqs.~(\ref{muqa})-(\ref{muqa2}).
The $\sigma$ meson field  is determined through
\begin{eqnarray}
\frac{\partial P}{\partial \sigma}=
\left( \frac{\partial P}{\partial M^{*}_{N}} \right)_{\mu_i,T}
\frac{\partial M^{*}_{N}}{\partial \sigma}
+\left(\frac{\partial P}{\partial \sigma}\right)_{M^{*}_{N}}=0.
\label{preseg}
\end{eqnarray}
\begin{equation}
m_\omega^2\omega_0 = \sum_{i=B} g_{\omega B}\rho_i ~,
\label{field2}
\end{equation}
\begin{equation}
m_\rho^2\rho_{03} = \sum_{i=B} g_{\rho B} I_{3B} \rho_i~.
\label{field3}
\end{equation}
The hyperon couplings are not
relevant to the ground state properties of nuclear matter, but information
about them can be available from the levels in $\Lambda$ hypernuclei
\cite{chrien,moszk,glen}:

$$g_{\sigma B}=x_{\sigma B}~ g_{\sigma N},~~g_{\omega B}=x_{\omega B}~
g_{\omega N}, ~~g_{\rho B}=x_{\rho B}~ g_{\rho N}$$
and $x_{\sigma B}$, $x_{\omega B}$ and $x_{\rho B}$ are equal to $1$ for the
nucleons and acquire different values in different parameterizations for the
other baryons. Note that the $s$-quark is unaffected by the sigma and omega
mesons i.e. $g_\sigma^s=g_\omega^s=0\ .$

For the bag radius  we take $R_N=0.6$ fm.
The two unknowns $Z_N$ and $B_N$ are obtained by fitting the 
nucleon mass $M=939$ MeV and
enforcing the stability condition for the bag at free space. The values
obtained are $Z_N=3.98699$ and $B_N^{1/4}=211.303$ MeV  for $m_u=m_d=0$ MeV and
$Z_N=4.00506$ and $B_N^{1/4}=210.854$ MeV for $m_u=m_d=5.5$ MeV.

Next we fit the quark-meson coupling constants $g_\sigma^q$,
$g_\omega = 3g_\omega^q$ and $g_\rho = g_\rho^q$ for the nucleon to obtain
the correct saturation properties of
the nuclear matter, $E_N \equiv   \epsilon/\rho - M = -15.7$~MeV at
$\rho~=\rho_0=~0.15$~fm$^{-3}$, $a_{sym}=32.5$ MeV, $K=257$ MeV and
$M^*=0.774 M$.

Moreover, as we are interested in obtaining also the production of pions and
kaons, they are introduced through Bose-Einstein distribution functions
\begin{equation}
\rho_i =
\frac{2J_M+1}{2\pi^2} \int_0^\infty p^2 dp
\left[\frac{1}{exp[(E_i-\mu_i)/T] -1} \right],
\label{rhopi}
\end{equation}
where $i=\pi^+,\pi^-,\pi^0, K^+, K^-,K^0, \bar K^0$, and the corresponding
vector mesons $\rho$ and $K^*$, with $J_M=0$ and 1.
$E_i=\sqrt{p^2 + m_i^2}$ and the chemical potentials are again written in
terms of their quark constituents, namely,
$\mu_{\pi^+}=\mu_u-\mu_d$, $\mu_{K^+}=\mu_u-\mu_s$ and so on. We have
considered that they behave like a free gas and their properties are not
changed due to their interaction with matter and, therefore, the fraction of 
produced
mesons is determined statistically from their free space properties.

In the sequel we compare the QMC results with the ones obtained within 
relativistic mean-field models NL3 \cite{nl3} and TW \cite{tw}. For 
reference we show in Table \ref{tab1} the saturation properties of the 
three models, all very similar.
\begin{table}
\caption{ Nuclear matter properties.}
\begin{center}
\begin{tabular}{cccc}
\hline
properties & NL3 & TW & QMC \\
& \cite{nl3} &\cite{tw} & \cite{qmc}\\
\hline
$B/A$ (MeV) & 16.3 & 16.3 &  15.7\\
$\rho_0$ (fm$^{-3}$) & 0.148 & 0.153 &0.150  \\
$K$ (MeV) & 271 & 240 & 258\\
${\cal E}_{sym.}$ (MeV) & 37.4 & 32.0 & 32.5  \\
$M^*/M$ & 0.60 & 0.56 & 0.77\\
\hline
\end{tabular}
\end{center}
\label{tab1}
\end{table}

In order to obtain the particle yields and respective densities three
conserved quantities are considered: the total strangeness is set to zero, the
total number of baryons and the total isospin are given by  
$N_B= 2 (N+Z)= 394$ and $I_3=(Z-N)/2=-39$ in a Au+Au collision and 
$N_B= 2 (N+Z)= 416$ and $I_3=(Z-N)/2=-44$ in a Pb-Pb collision.  
Our code deals with 6 unknowns, the three meson
fields and the three independent quark chemical potentials ($\mu_q, q=u,d,s$),
solved in a self-consistent manner.
We next analyze our results.
\section{Results}
\begin{figure}[th]
\includegraphics[width=0.9\linewidth]{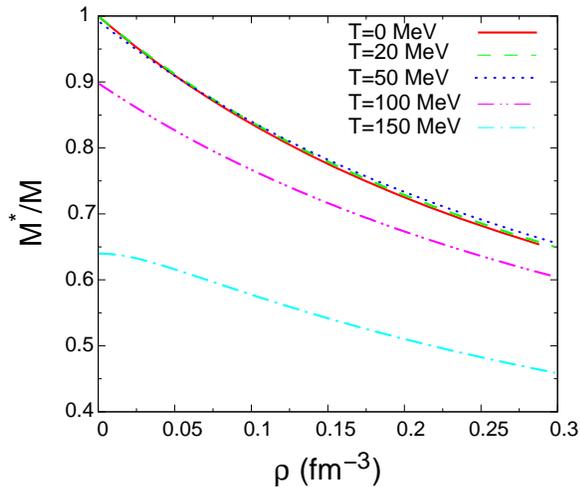}
\caption{Effective mass of the nucleon at finite temperature.}
\label{fig:effm}
\end{figure}
In Fig.\ref{fig:effm} the nucleon effective mass is displayed for different 
temperatures. As mentioned in the Introduction, one can see that the effective
mass decreases with the increase of the temperature for a fixed density
due to the identification of nucleon effective mass with the free 
energy of the bag.
Up to approximately 50 MeV the the effective mass does not vary much with
temperature but around 150 MeV, the temperature of interest for the present
study, the decrease with respect to T=0 is huge. This fact has direct
consequences in the values of the effective masses obtained for the
freeze-out temperature and baryonic densities as is discussed next.

\begin{figure}[t]
\includegraphics[width=0.9\linewidth]{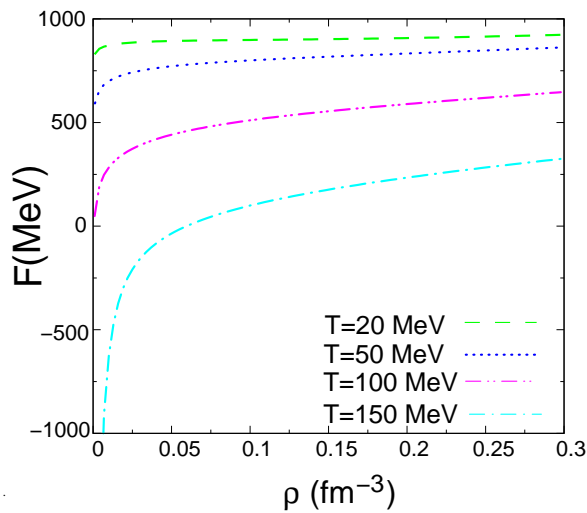}
\caption{Free energy of the nuclear matter at finite temperature.}
\label{fig:free}
\end{figure}

In Fig.\ref{fig:free} the nuclear matter free energy is shown for different
temperatures and it is seen that it increases considerably. As the free energy
of the bag is the main ingredient in the minimization procedure, understanding
its behaviour is important.

\begin{figure}[t]
\includegraphics[width=0.9\linewidth]{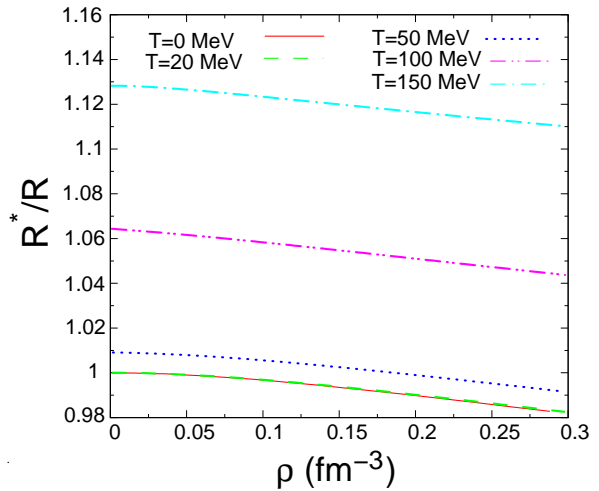}
\caption{Radius of the bag at medium at finite temperature.}
\label{fig:radius}
\end{figure}

In Fig.\ref{fig:radius} we show the bag radius for different temperatures and
we notice that it swells as the temperature increases.
This behaviour is contrary to the one which occurs if the effective mass
is identified with the energy of the bag, in which case the bag shrinks with
temperature. We expect that the bag swells with temperature and for a high
enough temperature the bag should dissolve. 

We finally consider the proposed calculation, i.e., the particle yields in
heavy ion collisions. For this purpose we have implemented a $\chi^2$ fit as 
in \cite{munzinger} to obtain the temperature and chemical potential of the
freeze-out:
\begin{equation}
\chi^2 = \sum_i \frac{({\cal R}_i^{exp} -{\cal R}_i^{theo})^2}
{\sigma_i^2},
\label{chi2}
\end{equation}
where ${\cal R}_i^{exp}$ and ${\cal R}_i^{theo}$ are the $i^{th}$ particle
ratio given experimentally and calculated with our models and $\sigma_i$
represents the errors in the experimental data points. If the description
of the data is consistent the minimum of $\chi^2$ should coincide with the
minimum of $q^2$ defined as
\begin{equation}
q^2 = \sum_i \frac{({\cal R}_i^{exp} -{\cal R}_i^{theo})^2}
{({\cal R}_i^{theo})^2}.
\label{q2}
\end{equation}
In obtaining the best fit values for the temperature and chemical potentials,
we have used the experimental ratios appearing in Table \ref{tab2} four times
for $\bar p/p$, twice for $\pi^-/\pi^+$ and four times for $K^-/K^+$, all with
the same weight. We have also taken into account the $K^{0*}/h^-$ and
$\bar K^{0*}/h^-$ ratios, where $h^-$ is the net sum of all negative
electrically charged hadrons. Instead we could have taken the mean value of
the measured values and a statistical average value of the errors.
\begin{table}
\caption{ Comparison of Au-Au experimental particle ratios (RHIC), 
relativistic mean field models and quark-meson coupling model results.}
\begin{center}
\begin{tabular}{cccccccccccc}
\hline
ratio & exp. data & exp & NL3  & TW & QMC\\
&&&  \cite{nl3} & \cite{tw} & \cite{qmc}\\
\hline
$\bar{p}/p$  & 0.65$\pm$0.07 & STAR & 0.650 & 0.656 & 0.738\\
             & 0.64$\pm$0.07 & PHENIX  & & \\
             & 0.60$\pm$0.07 & PHOBOS  & & \\
             & 0.64$\pm$0.07 & BRAHMS  & & \\
$\bar{p}/\pi^-$ & 0.08$\pm$0.01 & STAR &  0.075 & 0.076& 0.083 \\
$\pi^-/\pi^+$   & 1.00$\pm$0.02 & PHOBOS  & 0.998 & 1.01& 0.999 \\
                & 0.95$\pm$0.06 & BRAHMS & & &\\
$K^-/K^+$       & 0.88$\pm$0.05 & STAR & 0.912 & 0.896& 1.001 \\
                & 0.78$\pm$0.13 & PHENIX  &&&\\
                & 0.91$\pm$0.09 & PHOBOS  &&&\\
                & 0.89$\pm$0.07 & BRAHMS &&&\\
$K^-/\pi^-$     & 0.149$\pm$0.02& STAR & 0.234 & 0.228 & 0.191 \\
$\bar{\Lambda}/\Lambda$ & 0.77$\pm$0.07 & STAR & 0.681 &0.663 & 0.681 \\
$\bar \Xi^-/\Xi^-$ & 0.82$\pm$0.08 & STAR & 0.746 &0.739 &0.713  \\
$K^{0*}/h^-$ & 0.06 $\pm$ 0.017 & STAR & 0.058 & 0.064 & 0.037 \\
$\bar K^{0*}/h^-$ &0.058 $\pm$ 0.017 & STAR & 0.053 &0.056 &0.037 \\
\hline
\hline
$\Lambda/h^-$ && &0.021 & 0.023 &  0.023\\
\hline
$T$(MeV) &&&149 & 146.6 &132 \\
$\mu_b$ (MeV) &&& 47.5 & 62.8 &32.5  \\
$\rho$ $\times 10^{-3}$ (fm$^{-3}$) &&& 8.37 & 4.90 &3.08  \\
$\chi^2$ &&& 23.94 & 22.18 & 27.56 \\
$q^2$ &&& 0.21 & 0.19& 0.80\\
\hline
\end{tabular}
\end{center}
\label{tab2}
\end{table}
In Table \ref{tab2} we show the experimentally measured ratios, 
the QMC results calculated in the present work and the results obtained for
the  NL3 parametrization \cite{nl3} and the TW parametrization of a density
dependent hadron model \cite{tw}, given in \cite{nosso1}.
\begin{table}
\caption{Effective masses obtained with the temperature and chemical
potentials given in Table \ref{tab2}.}
\begin{center}
\begin{tabular}{cccccccccccc}
\hline
&  NL3  & TW & QMC \\
& \cite{nl3} & \cite{tw} & \cite{qmc} \\
\hline
$M^*_N/M_N$ & 0.88 & 0.87 &0.78 \\
$M^*_\Lambda/M_\Lambda$ & 0.93 & 0.92 &  0.83\\
$M^*_\Sigma/M_\Sigma$ & 0.94 & 0.93 & 0.84 \\
$M^*_\Xi/M_\Xi$ & 0.94 & 0.93 & 0.85\\
$M^*_{\Sigma^*}/M_{\Sigma^*}$ & 0.95 & 0.94 & 0.84 \\
$M^*_{\Xi^*}/M_{\Xi^*}$ & 0.94 & 0.94 & 0.72 \\
\hline
\end{tabular}
\end{center}
\label{tab3}
\end{table}
In Table \ref{tab3} we compare the baryon effective masses obtained within QMC
with the ones in NL3 and TW models for the temperature and baryon chemical
potential indicated in Table \ref{tab2}.
In this table $N$ stands for nucleons. It is seen that QMC predicts smaller
effective masses than NL3 and TW due to the internal structure of
baryons within this model, as discussed earlier. As a consequence in QMC 
we obtain larger values for the ratios antiparticles/particles, as well as 
$\bar p/\pi^-$, when compared with NL3 and TW. 
Smaller effective masses also allow the reproduction of the experimental 
ratios with a lower temperature. A lower temperature may explain the  
smaller kaon yields of QMC with respect to NL3 and TW, and therefore 
smaller ratios for $K^-/\pi^-$, and $K^{0*}/h^-$, $\bar K^{0*}/h^-$.

Anyhow, as in \cite{nosso1} the effective  masses and the value of the 
fireball temperature seem to show that the freeze out occurs below the 
critical temperature, at a temperature below the phase transition to a 
massless baryon phase.

QMC seems to do less well in reproducing the meson yields and this is probably
due to the simplified way mesons have been included in the present
calculation, just as free particles. Interactions would certainly affect their
effective masses and a better fit could probably be obtained as discussed in
\cite{nosso2}. In fact the mesons could also have been  described as bags
like in \cite{kaon}.

Next we  describe within QMC the Pb-Pb experimental particle ratios obtained at SPS supposing that chemical equilibrium was attained before freeze-out.
The authors of \cite{braun99} have shown within an improved statistical model 
with excluded volume and resonance decays, that the experimental hadronic 
yields and their ratios at SPS energy  are compatible with a  chemical
equilibrium population. From the whole set of existing experimental data we
have chosen all ratios involving the baryonic octet,  pions and charged
kaons. For the antiproton/proton ratio we took only the ones excluding feeding
from weak decays. Within QMC we were not able to fit the data with a $\chi^2$
lower than 216 and $q^2=17.3$. The results are given in Table \ref{tab4}, 
where the ratios obtained with NL3 are also shown. For NL3 the minimum 
$\chi^2$ is smaller but with a huge $q^2$. In the last column we give the NL3 
results for $q^2=17.3$ (NL3$^{\prime}$), the value obtained with QMC. For the  
new fit  the ratios get closer to the experimental data but the anti-particles 
and hyperons are still badly described.

We have not considered either heavier-resonance decays neither feeding from 
weak decays as done in \cite{braun99}. 
Weak decay affects mainly the ratios involving $\Xi$. If these rations are 
calculated as half the original ratio, for the NL3 model (shown in Table
\ref{tab4} as NL3*), the $\chi^2$ value improves considerably. But this is not 
the case when the QMC model is employed. It is important to mention that the 
best fit obtained for the SPS data in \cite{braun99} gave
$T=168 \pm 2.4$ and $\mu_b=266 \pm 5$, both values much higher than ours.
In \cite{braun99}, $\chi^2 = 37.8$ was also far away from the ideal value. 
\begin{table*}[t]
\caption{ Comparison of Pb-Pb experimental particle ratios (SPS), 
and quark-meson coupling model results. NL3$^*$ refers to a fit supposing that
 50\% of the $\Xi$ yield decays. NL3$^{'}$ corresponds to the fit with   
$q^2=17.3$, the value obtained with QMC.}
\begin{center}
\begin{tabular}{cccccccccccccccc}
\hline
ratio & exp. data & exp & QMC& NL3& NL3$^*$&NL3$^\prime$\\
&&&  \cite{qmc}&\\
\hline
$(p-\bar p)/h^-$ & 0.228 $\pm$ 0.029  & NA49 & 0.213&0.244&0.256&0.207\\
$\bar{p}/p$  & 0.055 $\pm$ 0.01 & NA44 & 0.095&0.0003&0.079& 0.003\\
$\Lambda/h^-$ & 0.077 $\pm$ 0.011 & WA97 & 0.007&0.00002 &0.008&0.002\\
$\pi^-/\pi^+$   & 1.1 $\pm$ 0.1 & NA49 & 0.997&1.066&1.101& 1.072\\
$K^+/K^-$       & 1.85 $\pm$ 0.09 & NA44 & 1.001& 1.854&1.824&1.656\\
                & 1.8 $\pm$ 0.1 & NA49 & &&\\
$\bar{\Lambda}/\Lambda$ & 0.131 $\pm$ 0.017 & WA97 & 0.087&0.0006&0.099&0.043\\
$ \Xi^-/\Lambda$ & 0.11 $\pm$ 0.01 & WA97 & 0.117&0.120&0.128&0.23\\
$\Xi^+/\bar \Lambda$ & 0.188 $\pm$ 0.039 & NA49 & 0.122&0.222&0.233&0.380\\
                          & 0.206 $\pm$ 0.04 & WA97 & &\\
$(\Xi^++\Xi^-)/(\bar{\Lambda}+\Lambda)$ & 0.13 $\pm$ 0.03 & NA49 & 0.117&0.120&0.138&0.237\\
$\Xi^+/\Xi^-$ & 0.232 $\pm$ 0.033 & NA49 &0.091&0.001&0.181 &0.071\\
              & 0.247 $\pm$ 0.043& WA97 & \\
\hline
$T$(MeV) &&&130 &99& 156.1&140.\\
$\mu_b$ (MeV) &&&167.5&411&330.2&303.\\
$\rho$ $\times 10^{-3}$ (fm$^{-3}$) &&&0.019&0.007& 0.064&0.030\\
$\chi^2$ &&& 216&186&20.07&285\\
$q^2$ &&& 17.32& 1.7$\times 10^5$& 0.48&17.31\\
\hline
\end{tabular}
\end{center}
\label{tab4}
\end{table*}
\section{Conclusions}
In the present work we have described the particle yield ratios in 
Au+Au at  RHIC and in Pb+Pb at SPS 
assuming thermal and chemical equilibrium within the QMC model, 
to test this model in the high temperature and low density regime.
Within the QMC model, nuclear
matter is described as a system of non-overlapping MIT bags which interact
through the effective scalar and vector mean fields \cite{qmc}.
We have identified the free energy of the baryonic MIT bag with the 
effective mass of each baryon and verified that, at finite temperature,  
the effective mass of the nucleons decreases with temperature and that 
a swelling of the MIT bag occurs.
In previous studies the energy of the MIT bag had been identified with 
the effective mass and, at finite temperature, this gave rise to an 
increase of the effective nucleonic mass with temperature, contrary to 
what occurs in other relativistic mean-field nuclear models, and a 
resulting decrease of the bag radius with temperature.

We have applied the QMC model to the description of the particle yield 
ratios in Au+Au at RHIC with success for a temperature of 132 MeV and 
a baryonic chemical potential of 32.5 MeV. These values are lower than 
the values obtained within a description using NLWM or density dependent 
hadronic models \cite{nosso1} because QMC predicts a faster reduction of 
the effective masses with temperature and therefore, for a given temperature, 
a larger anti-particle production. An improvement of the present calculation 
would also take into account the bag structure of the pions and kaons. 
As a consequence the effective mass of these mesons would change with 
temperature contrary to what has been considered in the present calculation.

We have also applied the QMC to the description of the particle production  
ratios in Pb+Pb in SPS. If the data are taken without the assumption of 
weak decays, the description of the experimental yield ratio is not very good
within the NL3 model, but improves when this assumption is considered. 
Within the QMC models the results are always poor.    
\section*{ACKNOWLEDGMENTS}
This work was partially supported by CNPq(Brazil), the FCT/CAPES-2009
collaboration and by FEDER/FCT (Portugal)
under the projects CERN/FP/83505/2008 and PTDC/FIS/64707/2006.

\end{document}